\documentclass[10pt,a4paper,twoside, twocolumn]{article}
\pdfoutput=1

\usepackage{graphicx}

\usepackage[T1]{fontenc}
\usepackage{amsmath}
\usepackage{ulem}

\usepackage{caption}
\usepackage{color}
\usepackage{colortbl}
\definecolor{Blue_1}{RGB}{204,225,246}
\definecolor{Blue_2}{RGB}{233,242,246}

\makeatletter
\renewcommand{\section}{\@startsection{section}{2}{0cm}{-\baselineskip}
{0,5\baselineskip}{\normalsize\bfseries}}
\renewcommand{\subsection}{\@startsection{subsection}{3}{0cm}{-\baselineskip}
{0,5\baselineskip}{\normalsize\slshape}}
\makeatother

\begin{document}

\title{Transit Time and Charge Correlations of Single Photoelectron Events in R7081 Photomultiplier
Tubes}
\author{F. Kaether, C. Langbrandtner}
\date{
 \small \sl 
 Max-Planck-Institut f\"{u}r Kernphysik, \\ 
 Saupfercheckweg 1, D-69117 Heidelberg Germany. \\
 E-mail: {\tt Florian.Kaether@mpi-hd.mpg.de} \\ 
}

\twocolumn[
\begin{@twocolumnfalse}
\maketitle
\begin{abstract}
\noindent During the calibration phase of the photomultiplier tubes (PMT) for the Double Chooz experiment the PMT response to light with single photoelectron (SPE) intensity was analysed. With our setup we were able to measure the combined transit time and charge response of the PMT and therefore we could deconstruct and analyse all physical effects having an influence on the PMT signal. Based on this analysis charge and time correlated probability density functions were developed to include the PMT response in a Monte Carlo simulation. \\
\end{abstract}
\end{@twocolumnfalse}
]

\section{Introduction}
Double Chooz \cite{DoubleChooz-proposal,DoubleChooz-result} is a reactor antineutrino experiment for the search of neutrino oscillations and a non vanishing neutrino mixing angle $\theta_{13}$ with two identical detectors located at different distances from the nuclear power plant at Chooz, France. Each detector will be observed by 390 PMTs to record light pulses produced by neutrino induced interactions inside a gadolinium loaded liquid scintillator \cite{gdscint}. The reaction $\bar\nu_e + p \to e^+ + n $ leads to a coincident signature with $e^+$ annihilation and a delayed $n$-capture by gadolinium or hydrogen with an energy release of 8 MeV or 2.2 MeV, respectively. With a light yield of 6000 photons per MeV, a PMT coverage of 13\% and a PMT sensitivity of 25\% an average signal of approximately 0.5 photoelectrons per PMT and MeV is expected. Therefore a precise knowledge of the characteristics of single photoelectron (SPE) signals is absolutely necessary for the understanding of the Double Chooz 
detector behaviour. Extensive characterizations of all Double Chooz PMTs are described in \cite{pmtgerman, pmtjapan}. Modified versions of this PMT type were also used and tested for other neutrino experiments \cite{antares, icecube}. In this article we focus on a more detailed measurement of the PMT response concerning the charge and transit time characteristics of SPE events and their physical aspects. Furthermore, the obtained results were used to build probability density functions (PDFs) for the Monte Carlo simulations of the Double Chooz detectors and the data analysis.

\section{Experimental setup}

The hemispherical PMT R7081 (Hamamatsu Photonics K.K.) has a diameter of 10 inch and a bialkali cathode. A more detailed describtion of the experimental setup is given in \cite{pmtgerman}. During the measurements, the PMTs are placed inside a Faraday dark room for light and electromagnetic noise protection while the complete electronics and DAQ are located outside to avoid crosstalk. In addition, the PMTs are surrounded by a cylindrical mu-metal \cite{mumetal} to shield the terrestrial magnetic field. The presence of a magnetic field would influence the trajectories of the photoelectrons between photocathode and first dynode leading to a degradation of the single photon electron spectrum. The shielding ensures a high energy resolution with a peak to valley ratio of $\sim 4.5$. As light source we use an A.L.S. picosecond laser with a wavelength of 438 nm which is in the region of the PMTs highest sensitivity. The laser light is diffused and led by optical fibres to the front of the PMTs to create a complete illumination of the photocathodes. A possible time jitter induced by the optical fibres is expected to be negligible, as we made some tests without the fibres with no observable influence on the transit time distribution. To create SPE signals we use a very low light intensity whereby only about 10\% of the triggered light pulses cause PMT responses. While the number $n$ of created photoelectrons is Poisson-distributed $p(n) = \mu^n/n! \, \exp(-\mu)$, the 
probabilities $p(n>1)$ are very low if the mean value is $\mu \approx 0.1$.

\begin{figure}[t]
\includegraphics[width=\columnwidth]{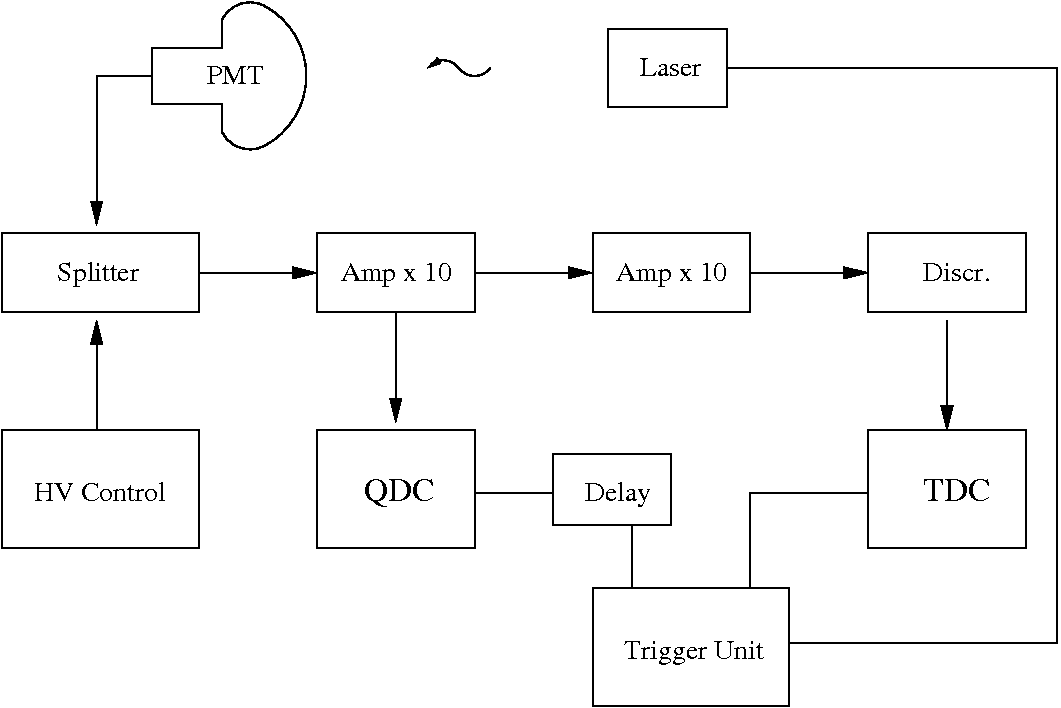}
\caption{Schematic overview of the experimental setup.}
\label{fig:setup}
\end{figure}

An overview on the electronical setup is shown in Figure \ref{fig:setup}. The Trigger system controls the laser output, defines the starting time for the Time to Digital Converter (TDC) and creates the time window for current integration with the Charge to Digital Converter (QDC). More details about the used electronical modules can be found in \cite{pmtgerman}.

The value of the high voltage was calibrated for a gain of $10^7$. While the PMTs are back terminated with 50$\,\Omega$ an average SPE charge of $0.5 \times 10^7 \, e = 0.8 \; \rm pC$ is expected. The PMT signals are decoupled from the high voltage with the splitter box. After  amplification the charge is determined and a discriminator forms the stop signal for the single-hit TDC. The VME based data acquisition allows to read out TDC and QDC simultaneously for each event. 

\section{Analysis of the Time and Charge Spectrum}

\begin{figure}[t]
\includegraphics[width=1\columnwidth]{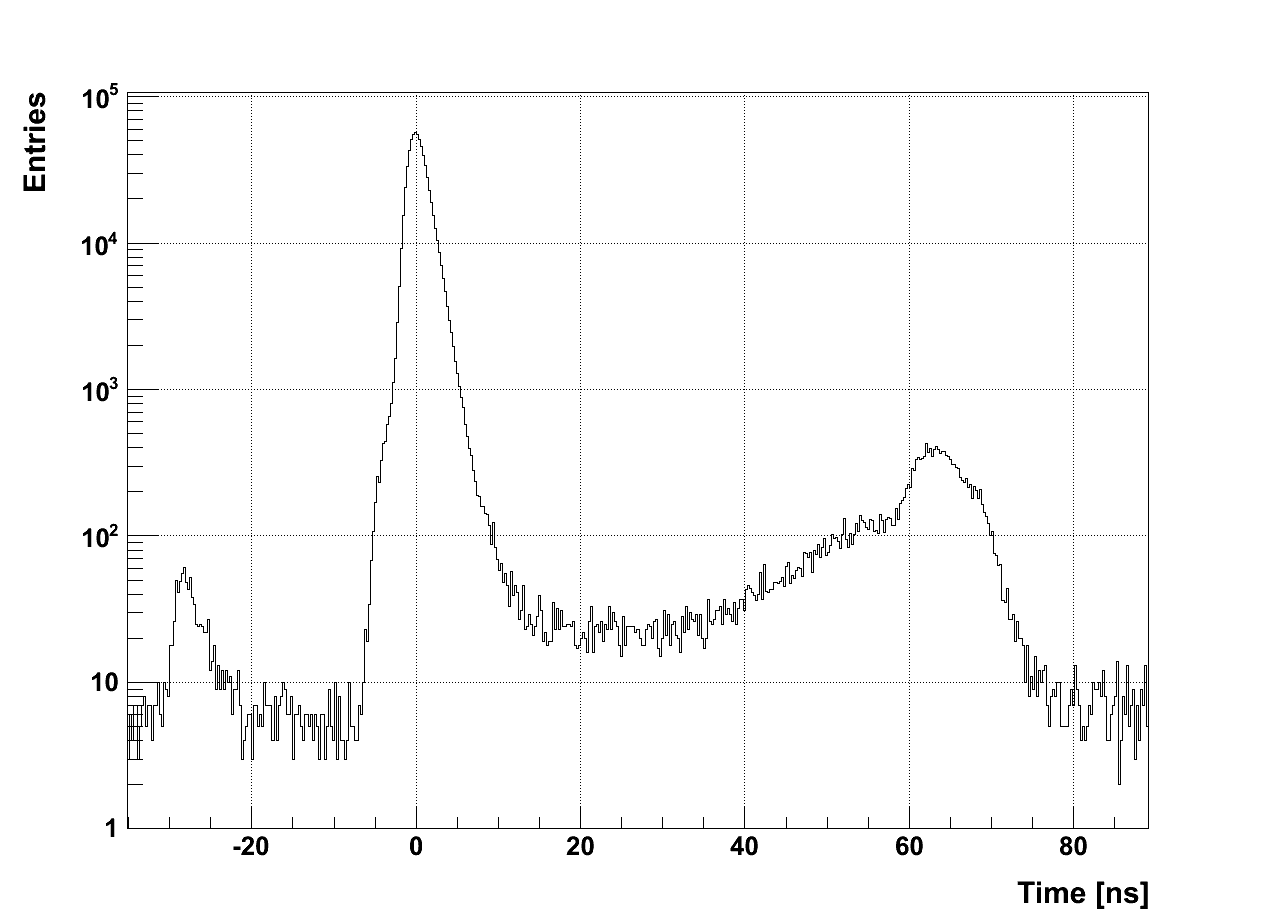}
\caption{Transit time spectrum of R7081 with SPE light intensity.}
\label{fig:TDC}
\end{figure}

\begin{figure}[t]
\includegraphics[width=1\columnwidth]{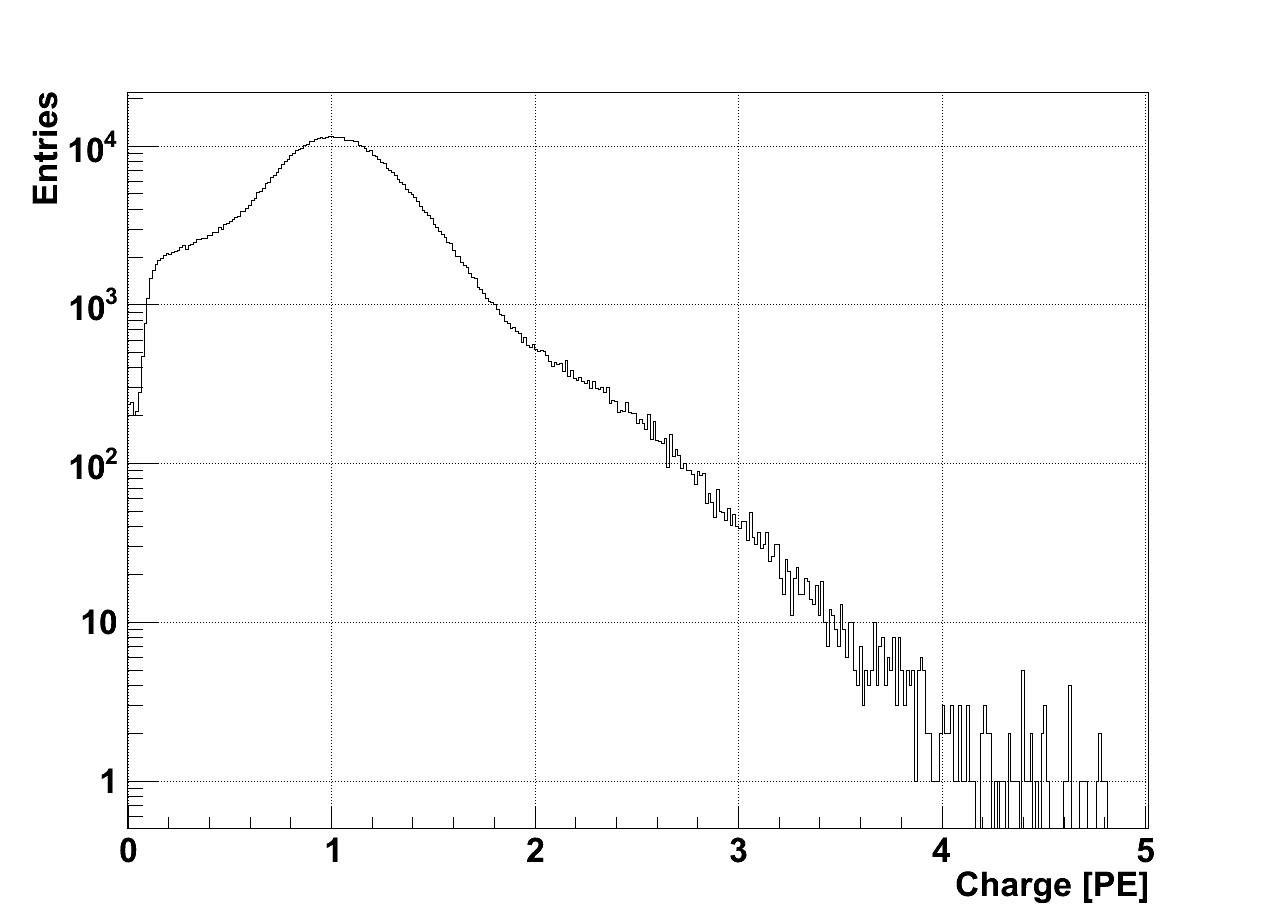}
\caption{Charge spectrum of R7081 with SPE light intensity.}
\label{fig:QDC}
\end{figure}

In this section we describe the effects occuring during the signal generation in the photomultiplier tubes resulting in  the observed structures in the transit time and charge distributions.

Figure \ref{fig:TDC} and \ref{fig:QDC} show the transit time and charge spectrum of the R7081 photomultiplier using laser light with SPE intensity. Since the absolute value of the transit time is biased by cable lengths, proceeding electronics etc. we focus on relative times while the maximum of the distribution is arbitrarily defined as $t=0$. Regarding the relative transit times in Figure \ref{fig:TDC}, three types of events can be classified:
\begin{list}{$\bullet$}{
  \setlength{\topsep}{1ex}
  \setlength{\itemsep}{-0.3ex}
  \setlength{\leftmargin}{3.5ex}
}
\item Main peak pulses: the major part of all PMT pulses (97$\,$\%) are located in the time interval $-10\,{\rm ns} \, < t < 10\,{\rm ns}$ including mostly regularly multiplied electrons.
\item Pre-pulses: roughly 30$\,$ns prior to the main peak so called pre-pulses occur, which arise from  photoelectrons directly produced at the first dynode.
\item Late-pulses appear up to 70$\,$ns after the main peak. They consist of  photoelectrons which are (in-)elastically back scattered from the first dynode and produce secondary electrons after being reaccelerated to the dynode structure.
\end{list}
\begin{figure}
\includegraphics[width=1\columnwidth]{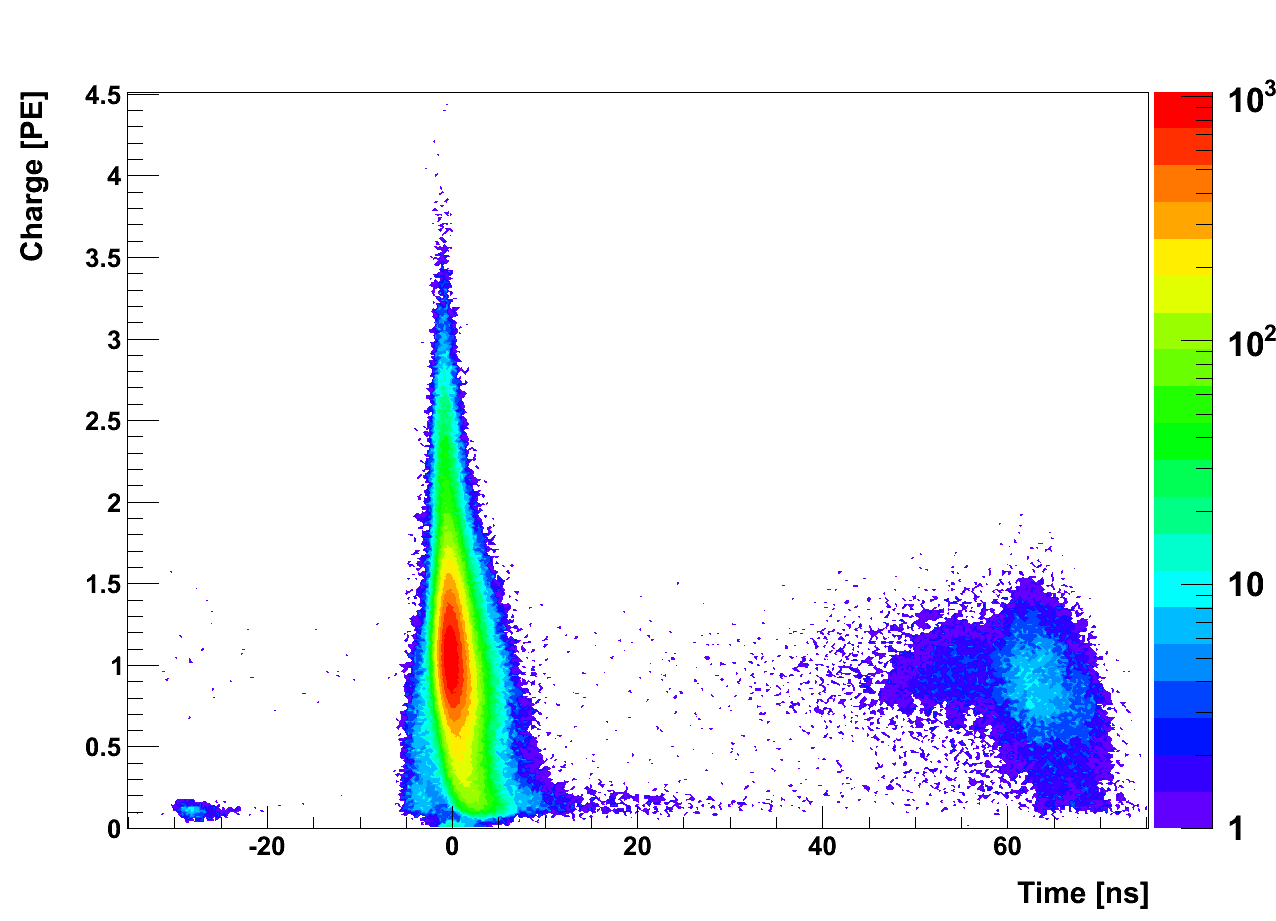}
\caption{Charge versus transit time distribution at SPE intensity.}
\label{fig:QDC_TDC}
\end{figure}
Since the dark noise rate of the PMTs (caused by thermal emission of electrons from the cathode) is in the range of a few thousand events per second, there is an observable accidental background in Figure \ref{fig:TDC} in form of a flat distribution with approximately 5 events per bin.

The charge of all events versus the transit time is shown in Figure \ref{fig:QDC_TDC}. In the following sections we will investigate the different distributions and their physical origin in more detail.

\subsection{Main peak pulses}

\begin{figure}[t]
\includegraphics[width=1\columnwidth]{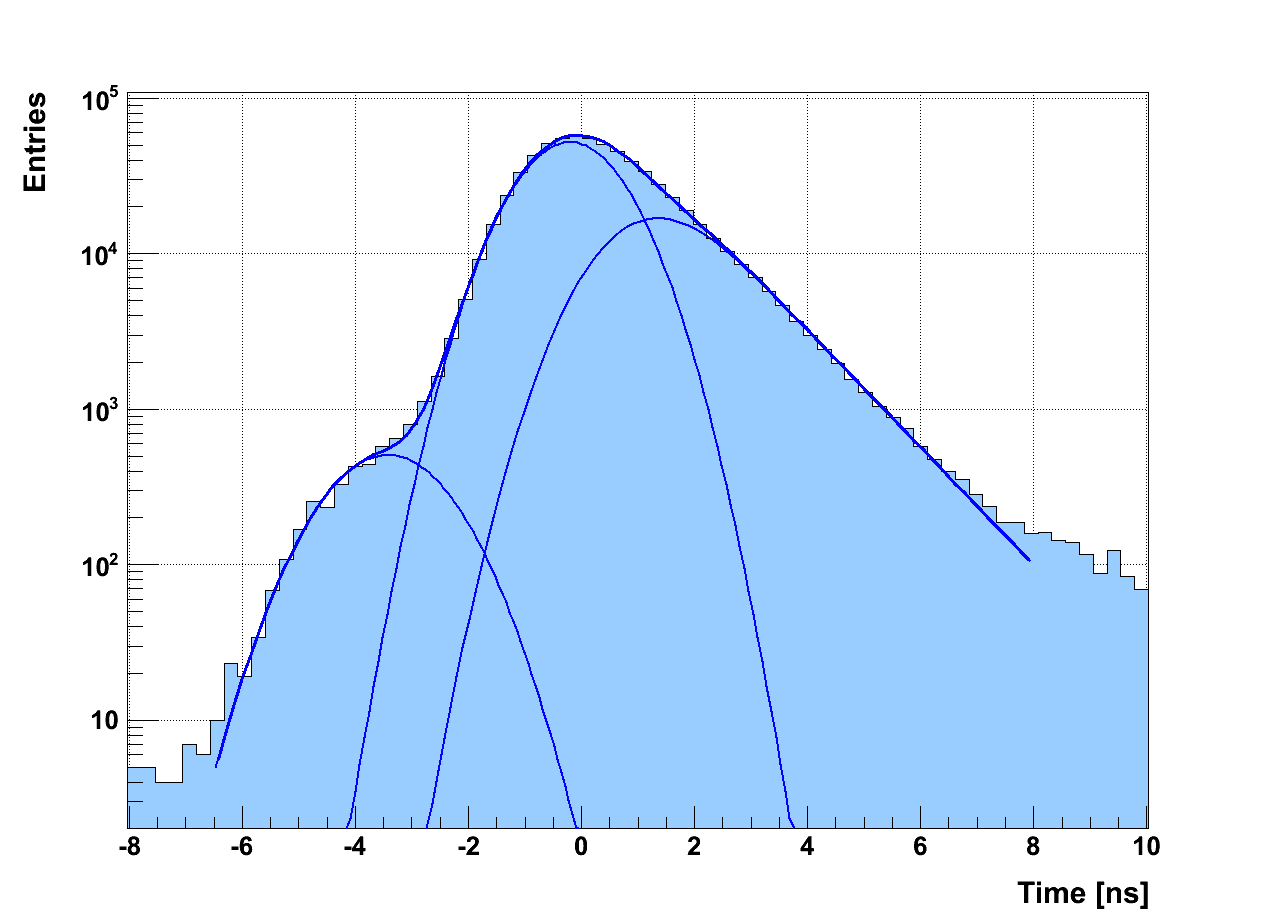}
\caption{Transit time of the Main peak fitted with two Gaussian curves and one expontential distribution  convoluted with a Gaussian with the same $\sigma$ as the central Gaussian.}
\label{fig:TT_Main_Peak}
\end{figure}
\begin{figure}[t]
\includegraphics[width=1\columnwidth]{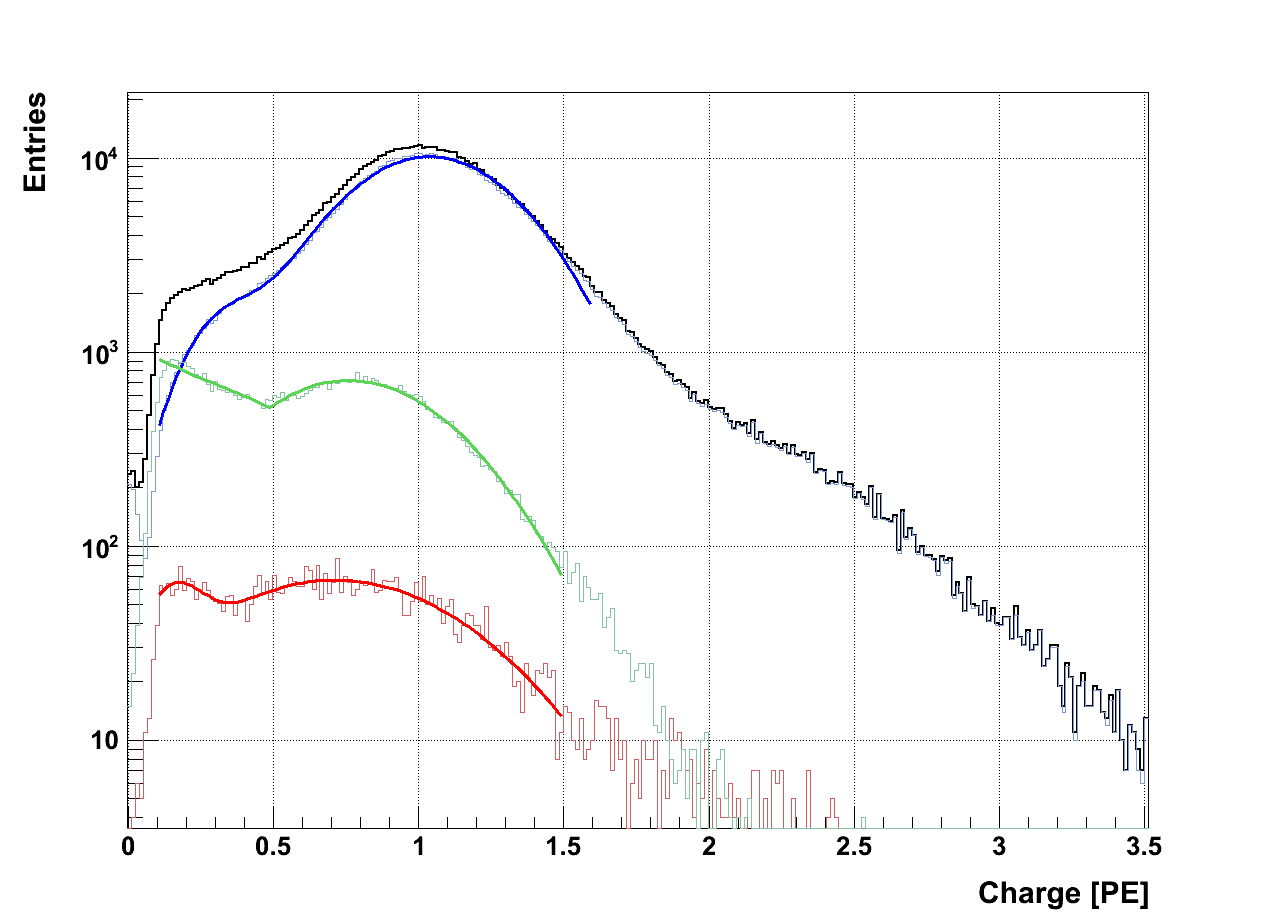}
\caption{
Charge distribution of the main peak pulses. 
{\it \bf Black:} Total charge spectrum. 
{\it \bf Blue:} Regular pulses in a range of $-2.5 \, {\rm ns} <  t  < 2.5\,{\rm ns}$ fitted by two Gaussian functions. 
{\it \bf Green:} Delayed pulses in the range of $2.5 \, {\rm ns} <  t  < 8 \,{\rm ns}$  fitted with a linear and a Gaussian functions.
{\it \bf Red:} Early pulses in the range of $-6 \, {\rm ns} <  t  < -2.5\,{\rm ns}$  fitted with two Gaussian functions.}
\label{fig:Charge_Main_Peak}
\end{figure}

Figure \ref{fig:TT_Main_Peak} zooms into the time distribution of the main peak in the time interval of $-8 \, {\rm ns} < t < 10 \, {\rm ns}$. The observed structures can be described by the sum of two Gaussians (early and regular pulses, respectively) plus a convolution of an exponential distribution with a Gaussian (delayed pulses). 
In Figure \ref{fig:Charge_Main_Peak} the charge distributions of the three different types of pulses are given, while their physical reasons are  discussed in the following.

\textit{Early pulses:}
The Gaussian distribution in the rising part of the main peak (see Figure \ref{fig:TT_Main_Peak}) with a mean of $t_E = -3.4 \, {\rm ns}$ and $\sigma_E = 1.0 \, {\rm ns}$ includes the so called early pulses. Their origin can be explained by an elastic forward scattering of photoelectrons at the first dynode. These photoelectrons keep their energy and directly hit the second dynode a few nanoseconds earlier than usual \cite{Borexino}. The first multiplication process creates the same factor as usual, but happens at the second dynode. Therefore one would expect the charge of these pulses to be reduced by the amplification factor of the second dynode which is estimated to be around 5. From the fit of these pulses with two Gaussians (see Figure \ref{fig:Charge_Main_Peak}) we get for the smaller one a mean charge of 0.15$\pm0.02\,$PE which is in quite good agreement with an expected value of 0.18$\,$PE. The second Gaussian around 0.7$\,$PE might be explained by inelastic forward scattered photoelectrons at the first dynode.

\textit{Regular pulses:}
The central Gaussian with the mean at  $t_R = -0.2 \, {\rm ns}$ and the so called transit time spread of $\sigma_R = 0.9 \, {\rm ns}$  consists of the predominant regular accelerated electrons. The transit time spread $\sigma_R$ is caused by different trajectories of the photoelectrons through the accelerating electrical field mainly depending on their starting positions and velocities at the cathode. In contrary, during the multiplication process in the dynode system, a time-dispersion results in pulse broadening rather than in transit time spread because of the large number of secondary electrons \cite{becker}. In Figure \ref{fig:Charge_Main_Peak} the charge of the regular electrons is shown and described by a Gaussian around 1.03$\,$PE with a sigma of 0.3$\,$PE.  In addition, there is a small fraction of not optimally multiplicated electrons falling into the applied time cut described by a second Gaussian at 0.3$\,$PE and a sigma of 0.14$\,$PE. The relative probability ratio of both Gaussians is 20:1.

\textit{Delayed pulses:}
The time distribution of the pulses in the falling part of the main peak is well described by a convolution of an exponential distribution and a Gaussian (see Figure \ref{fig:TT_Main_Peak}). For the Gaussian $\sigma$ we expect the same underlying physics as for the transit time spread, therefore we choose $\sigma_D = \sigma_R$. The delay of these events can be explained by photoelectrons produced at the outer side of the cathode running through a suboptimal acceleration due to field inhomogeneities between outer cathode and the first dynode. Moreover, the facing direction of the first dynode is possibly able to generate asymmetric electron paths. Both processes will cause not only a delay but also a lower multiplication at the first dynode, which can be seen in the charge distribution in Figure \ref{fig:Charge_Main_Peak}, where the charge fit is composed of a gaussian and an exponential (green). 

\subsection{Pre-pulses}

\begin{figure}[t]
\includegraphics[width=1\columnwidth]{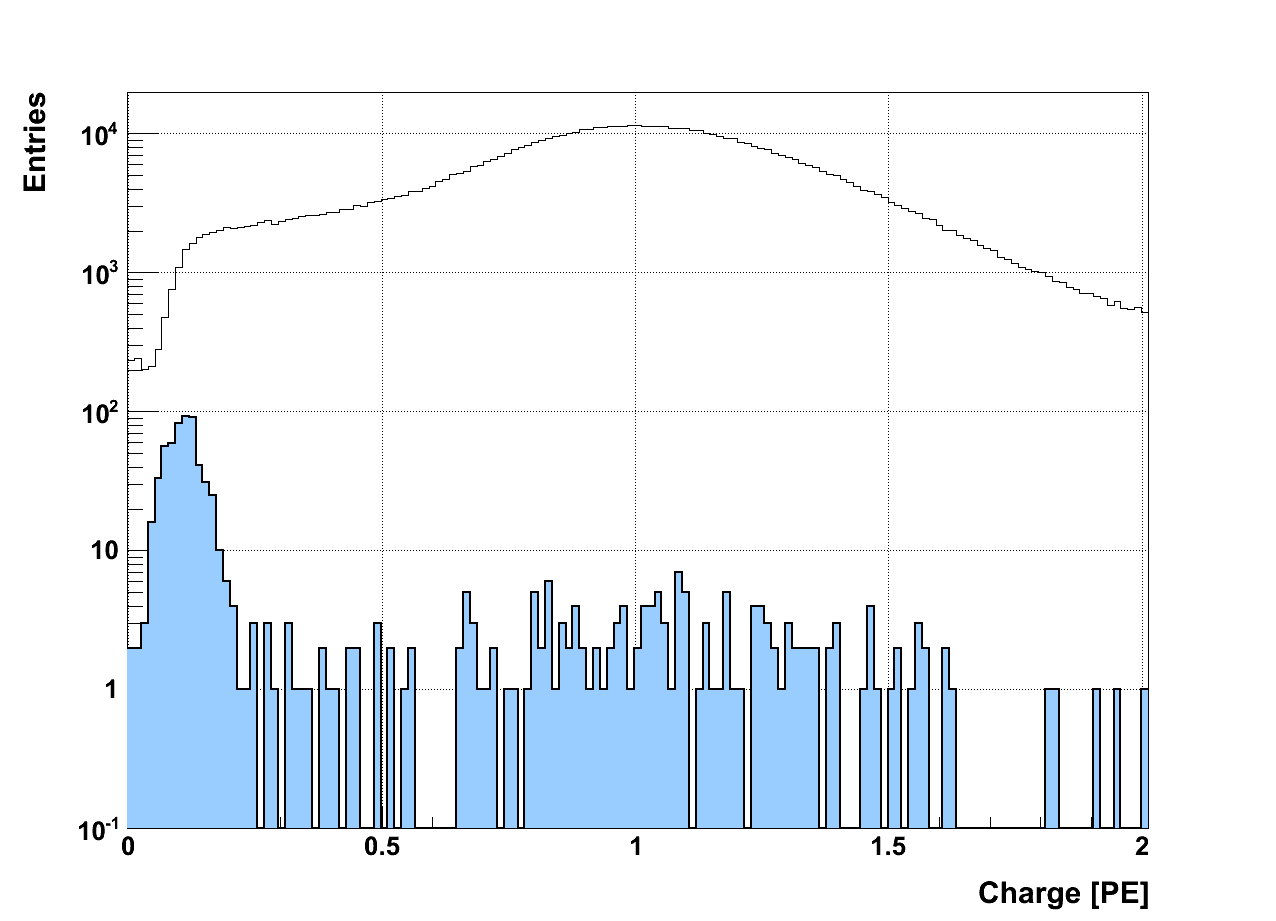}
\caption{Charge distribution of the pre-pulses (blue) and whole spectrum (black).}
\label{fig:Charge_Pre_Pulses}
\end{figure}

Pre-pulses occur if the photon passes through the cathode and directly hits the first dynode. The difference of the transit time with respect to regular pulses of roughly 30 ns is caused by the difference of the electron time of flight in the accelerating field between cathode and first dynode compared to the speed of light.  The charge of the pre-pulses is reduced by the amplification factor of the first dynode which is expected to be  around 15 for the R7081 series. Since the discriminator threshold for the PMT signal was 0.1$\,$PE  in our setup, we could only measure the higher part of the pre-pulse charge spectrum (Figure \ref{fig:Charge_Pre_Pulses}). For this part we determined a pre-pulse probability of 0.1$\,$\%, but this number also depends on the position of the light source and the angular distribution of the illumination. The events with charge values around 1$\,$PE in figure \ref{fig:Charge_Pre_Pulses} are accidental dark noise signals.

\subsection{Late-pulses}

\begin{figure}[t]
\includegraphics[width=1\columnwidth]{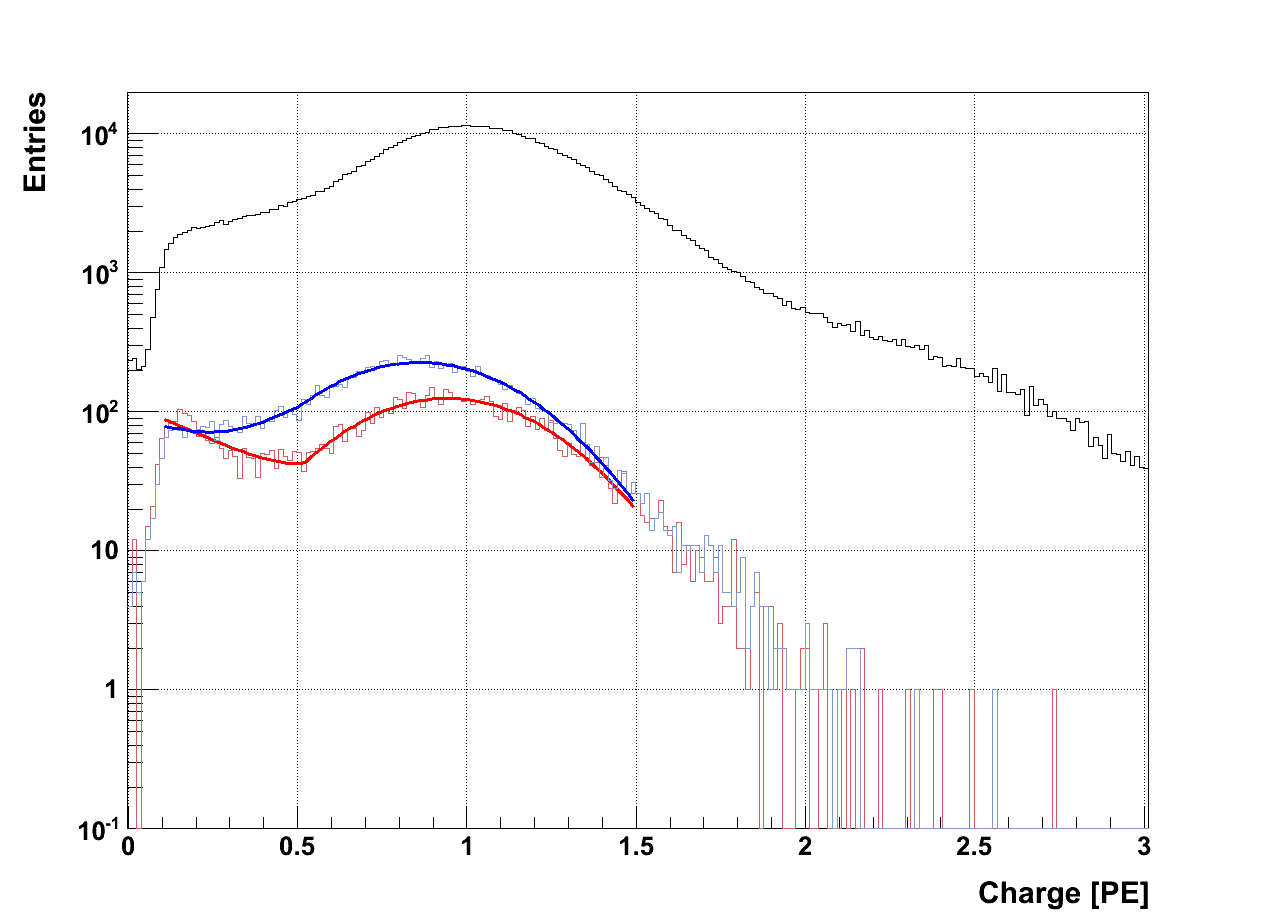}
\caption{Charge distribution of the late-pulses.
{\it \bf Black:} total charge spectrum.
{\it \bf Blue:} charge spectrum of mostly elastic scattered late-pulses with 60$\,$ns<$\,$t$\,$<75$\,$ns.
{\it \bf Red:} charge spectrum of inelastic late-pulses with 15$\,$ns<$\,$t$\,$<60$\,$ns, both fitted by a linear and a Gaussian function.}
\label{fig:Charge_Late_Pulses}
\end{figure}

Immediately after the main peak and up to 70$\,$ns so called late-pulses occur. Since the nomenclature found in literature is inconsistent, we want to clearly distinguish late-pulses from after-pulses. The latter are created by the ionisation of residual gas in the photomultiplier by photoelectrons. After-pulses have typical arrival times of some $\mu$s after the ionisation process and are therefore always correlated to a previous PMT signal. Since the TDC measures only the next signal after trigger-start, after-pulses do not influence the topic of this paper and are not further discussed. A description of the after-pulse behaviour of the PMT R7081 can be found in \cite{afterpulse}.

Late-pulses occur if the photoelectron is back scattered from the first dynode, then reaccelerated by the electrical field and finally produces secondary electrons hitting the first dynode for the second time. This process of back scattering can happen elastic or inelastic producing heat at the dynode  \cite{Lubsandorzhiev}. This explanation is consistent with the observation that the arrival time of late-pulses ends at around 70$\,$ns after the regular pulses which is approximately twice the time of flight of the photoelectrons  between the cathode and the first dynode ($\equiv$ pre-pulse time). It is slightly longer since the path length of the back scattered and back accelerated electron is not a straight line but rather a curve in most cases.

To distinguish between inelastically and elastically scattered photoelectrons we separate the late-pulses by their arrival times before and after 60 ns, respectively. This separation is motivated by the expectation of shorter path lengths of inelastically scattered photoelectrons. The charge distibutions of both samples are shown in Figure \ref{fig:Charge_Late_Pulses}. The separation of (in-)elastic events does not work perfectly well, however the fraction of events with lower energies is higher at earlier times, as expected. The charge distributions are described by a linear and a Gaussian part, while the origin of the linear distribution is expected to be caused by inelastically scattered electrons. The Gaussian charge distribution is expected to be generated by elastically scattered photoelectrons.

In our setup we measured the probability for late-pulses after t>10$\,$ns to be 3$\,$\%. In the description of the Monte Carlo simulation given in the next section we will see that an elastic to inelastic late-pulse ratio of 1:2 fits the data best.

\section{Simulation of the PMT response}

\subsection{Probability density functions}

\begin{table}[t]
\footnotesize
\centering
\begin{tabular}{|l c c c|}\hline
\rowcolor{Blue_1}Pulse & Time & Charge & Prob. $\left[\%\right]$ \\
Pre-pulses & G & G & 0.1\\
\rowcolor{Blue_2}Early-pulses & G & 2$\,$G & 1\\
Regular pulses & G & 2$\,$G & 76.9\\
\rowcolor{Blue_2}Falling edge pulses &  ${\rm G} \otimes \exp$ & $\rm G+lin $ & 19\\
Inelastic late-pulses & ${\rm G} \otimes \exp$ & $\rm G+lin $ & 2\\
\rowcolor{Blue_2}Elastic late-pulses & G & $\rm G+lin $ & 1\\\hline
\end{tabular}
\caption{PDFs for the different pulses. G stands for a Gaussian, lin for a linear and exp for an exponential probability distribution. \label{tab:PDFs}}
\end{table}

\begin{figure*}[!p]
\includegraphics[width=1\columnwidth]{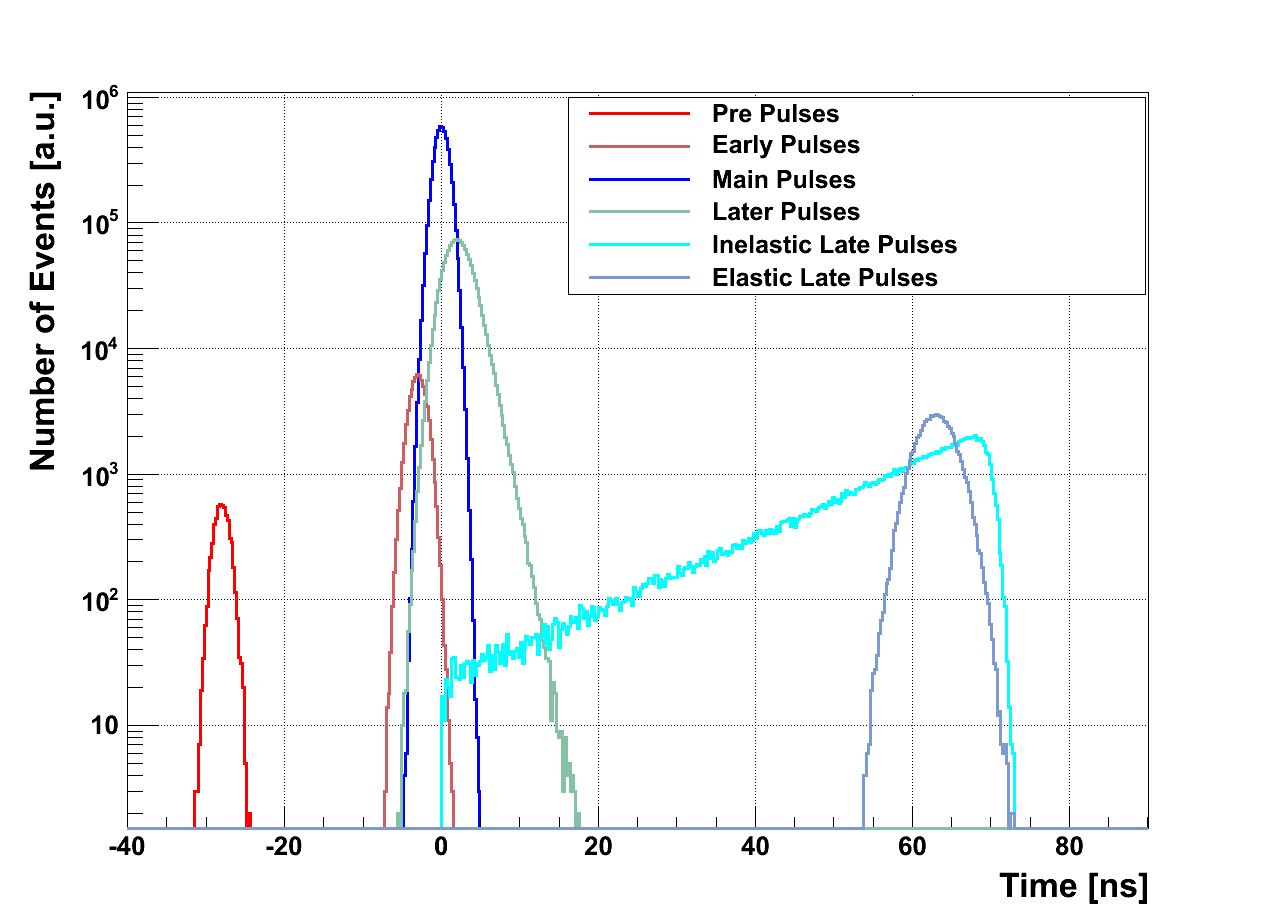}
\includegraphics[width=1\columnwidth]{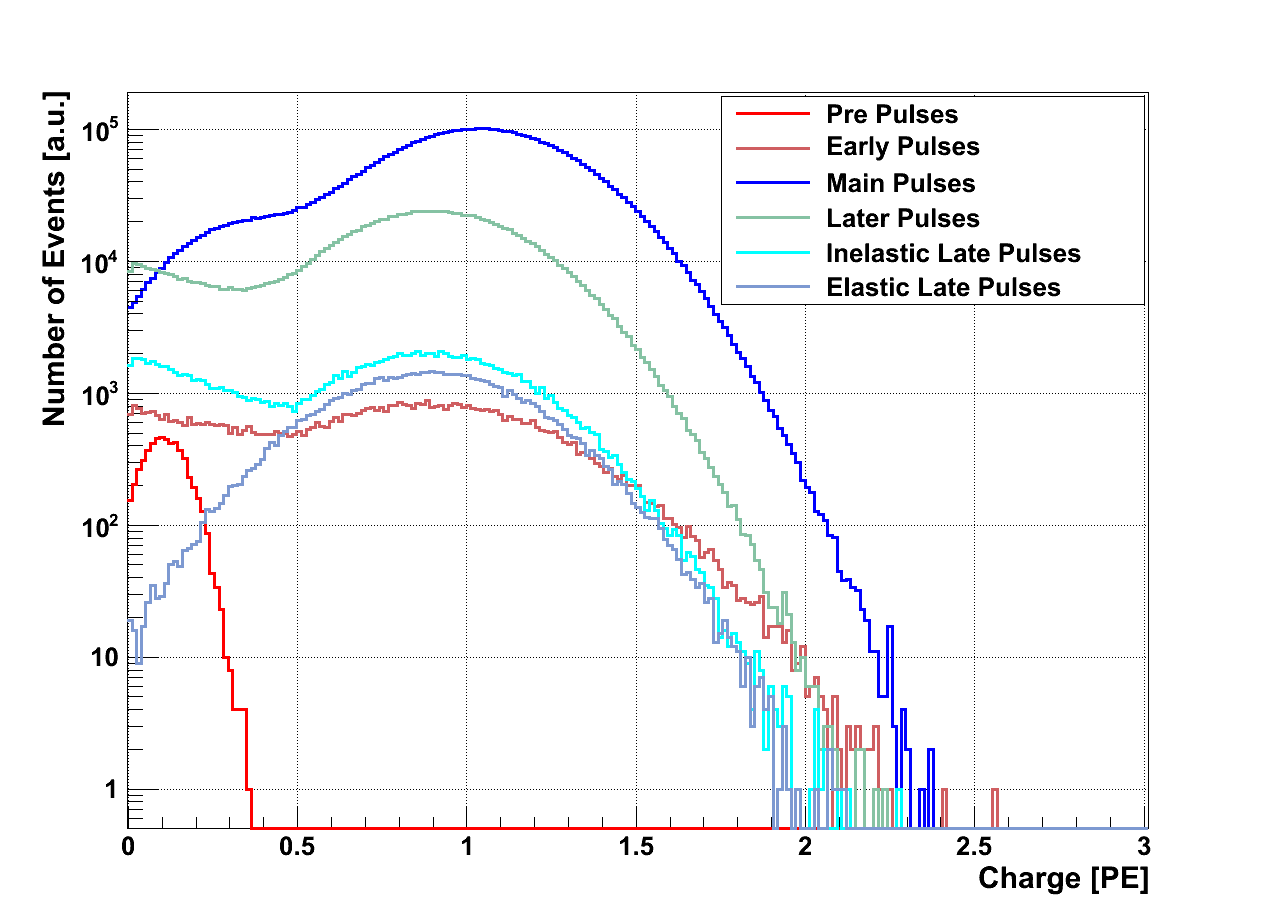}
\caption{Simulated transit time (left) and charge (right) compositions.}
\label{fig:Simulated_charge+time_Composition}
\end{figure*}

In the Double Chooz experiment the PMT charge and time response influences not only the energy reconstruction of detector events (e.g. from neutrino interactions), but also higher order analysis tools as vertex reconstruction and pulseshape analysis. Therefore it is crucial to understand the transit time and charge behaviour of the PMTs and to include them into the detector simulation in an appropriate way. In the previous section it was shown that effects inside the PMT having an influence on charge and time response can be well described by six different types of pulses, namely one type for pre-pulses, three types for main peak pulses and two types for late-pulses. The measurements described in the previous sections are the basis for probability density functions (PDFs) which are derived from the data fits for all these types of pulses and are listed in Table \ref{tab:PDFs}. Their composition in transit time and charge is given in Figure \ref{fig:Simulated_charge+time_Composition}.


\begin{figure*}[!p]
\includegraphics[width=1\columnwidth]{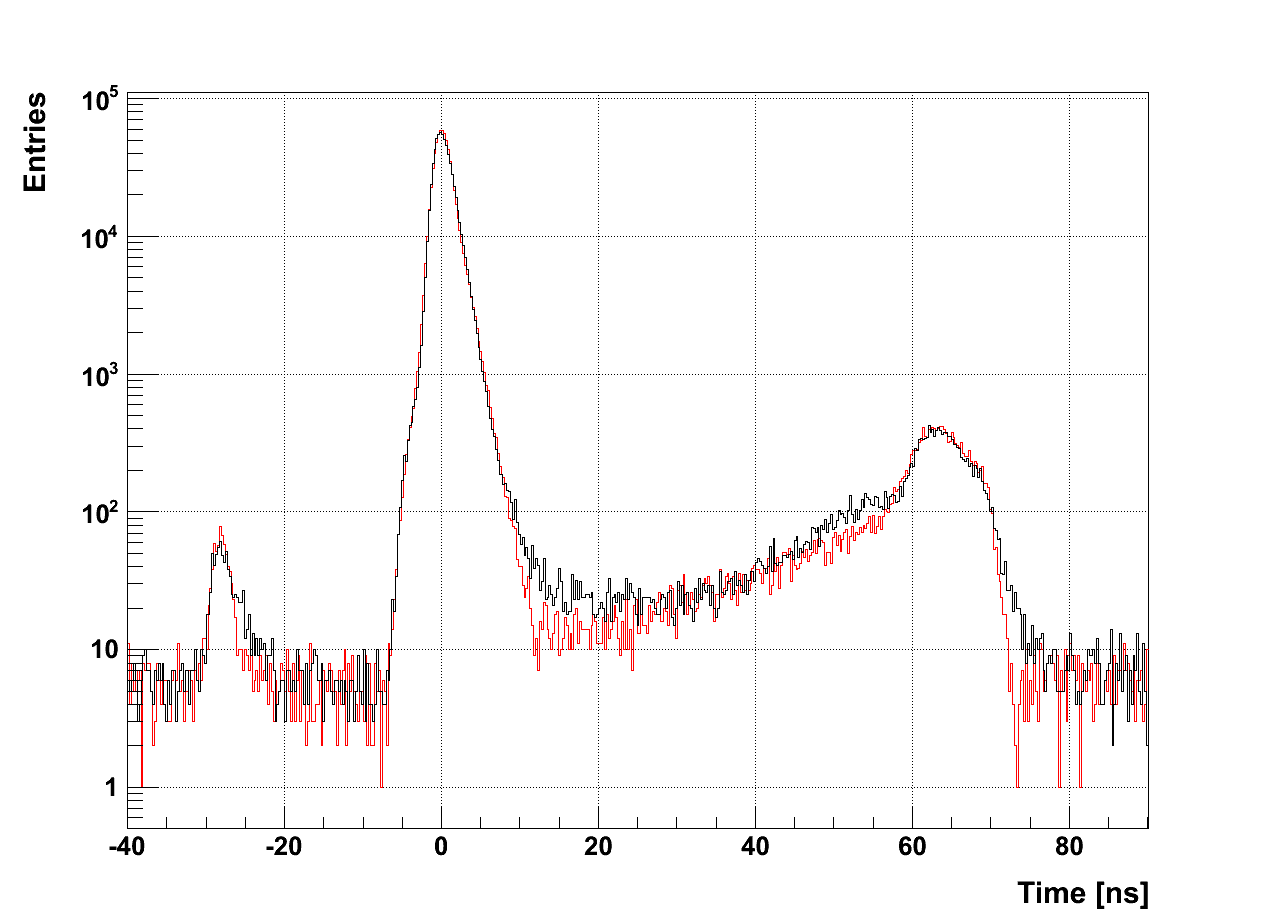}
\includegraphics[width=1\columnwidth]{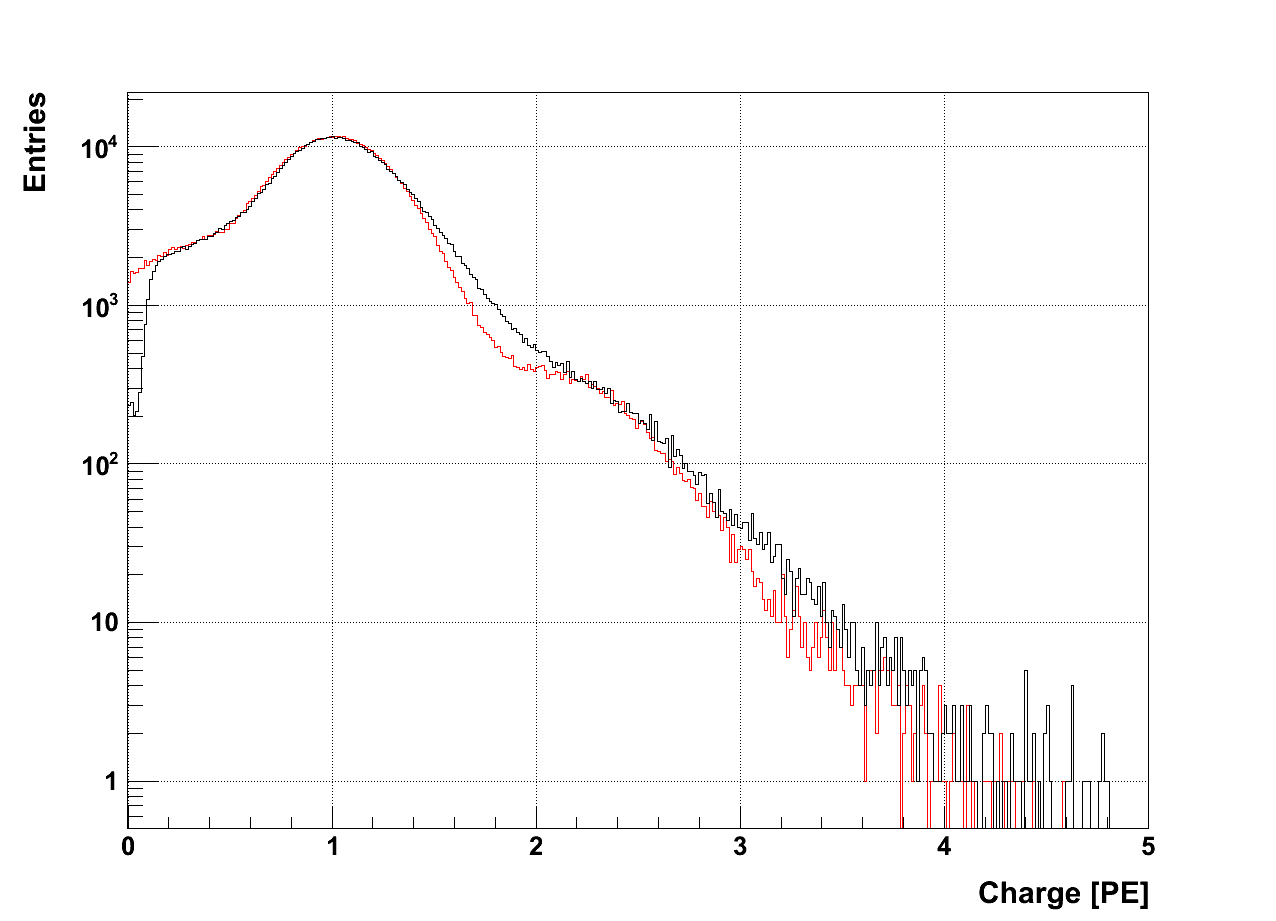}
\caption{Simulated (red) and real (black) transit time (left) and charge (right) spectra. For a better comparison uniformly distributed background signals were added to the Monte Carlo spectrum.}
\label{fig:Simulated_real_time+charge_spectrum}
\end{figure*}
\begin{figure*}[!p]
\includegraphics[width=1\columnwidth]{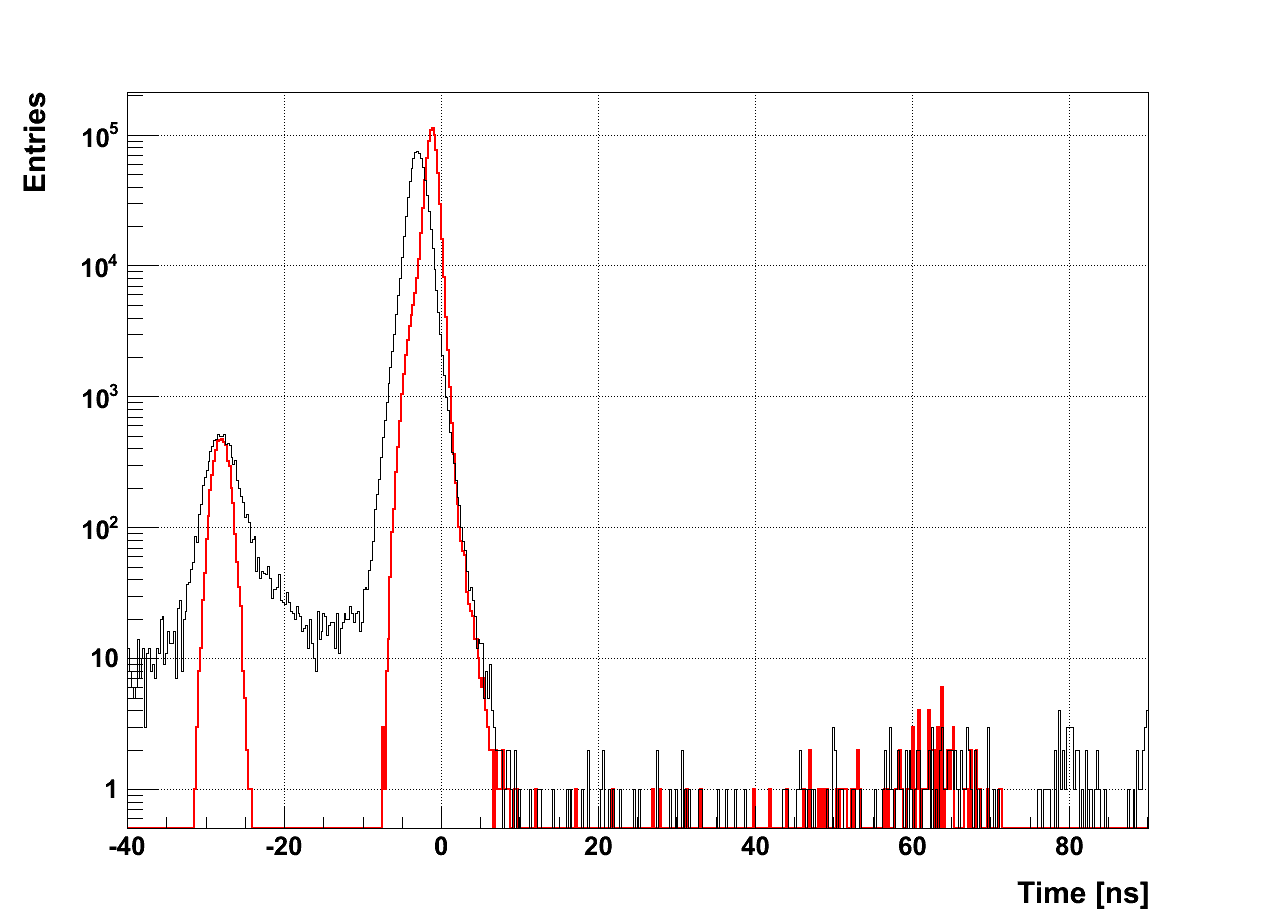}
\includegraphics[width=1\columnwidth]{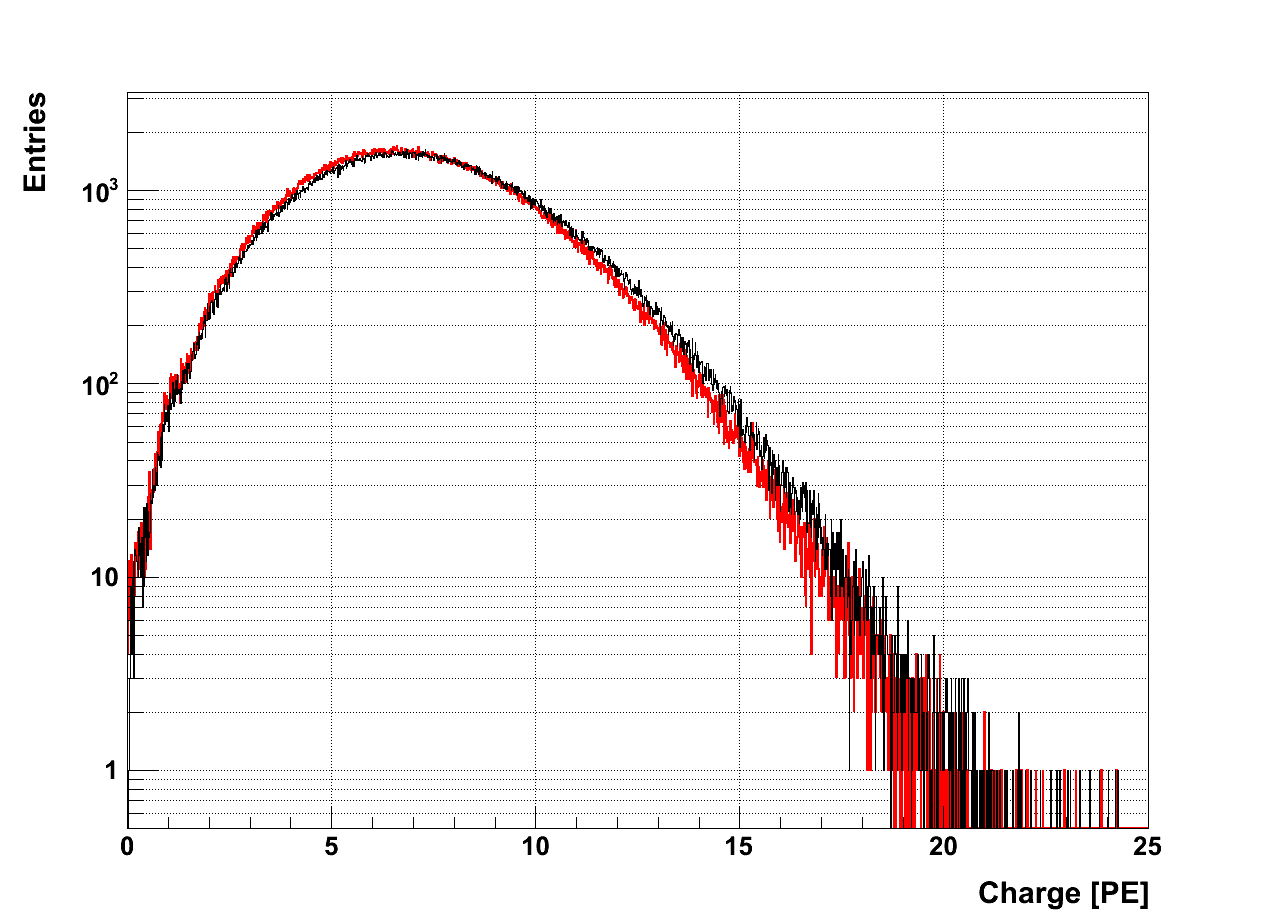}
\caption{Simulated (red) and measured (black) time (left) and charge (right) distribution for a multi PE light intensity with a Poissonian mean of $\mu=7.7$.}
 \label{fig:Simulated_real_time+charge_spectrum_high_int}
\end{figure*}

\subsection{Monte Carlo Simulation result}

The steps of the Monte Carlo Simulation are as follows: first it is decided by a Poisson distribution how many photoelectrons are created at the cathode. The next step decides which type of pulse from Table \ref{tab:PDFs} occurs and according to the PDFs for this pulse type the charge and transit time is chosen. This step is done for each photoelectron separately. The last step calculates the sum of all occurring photoelectrons which finally gives the total charge of the event while the transit time is calculated by the shortest arrival time of all pulses in the event. 

\begin{figure}[t]
\includegraphics[width=1\columnwidth]{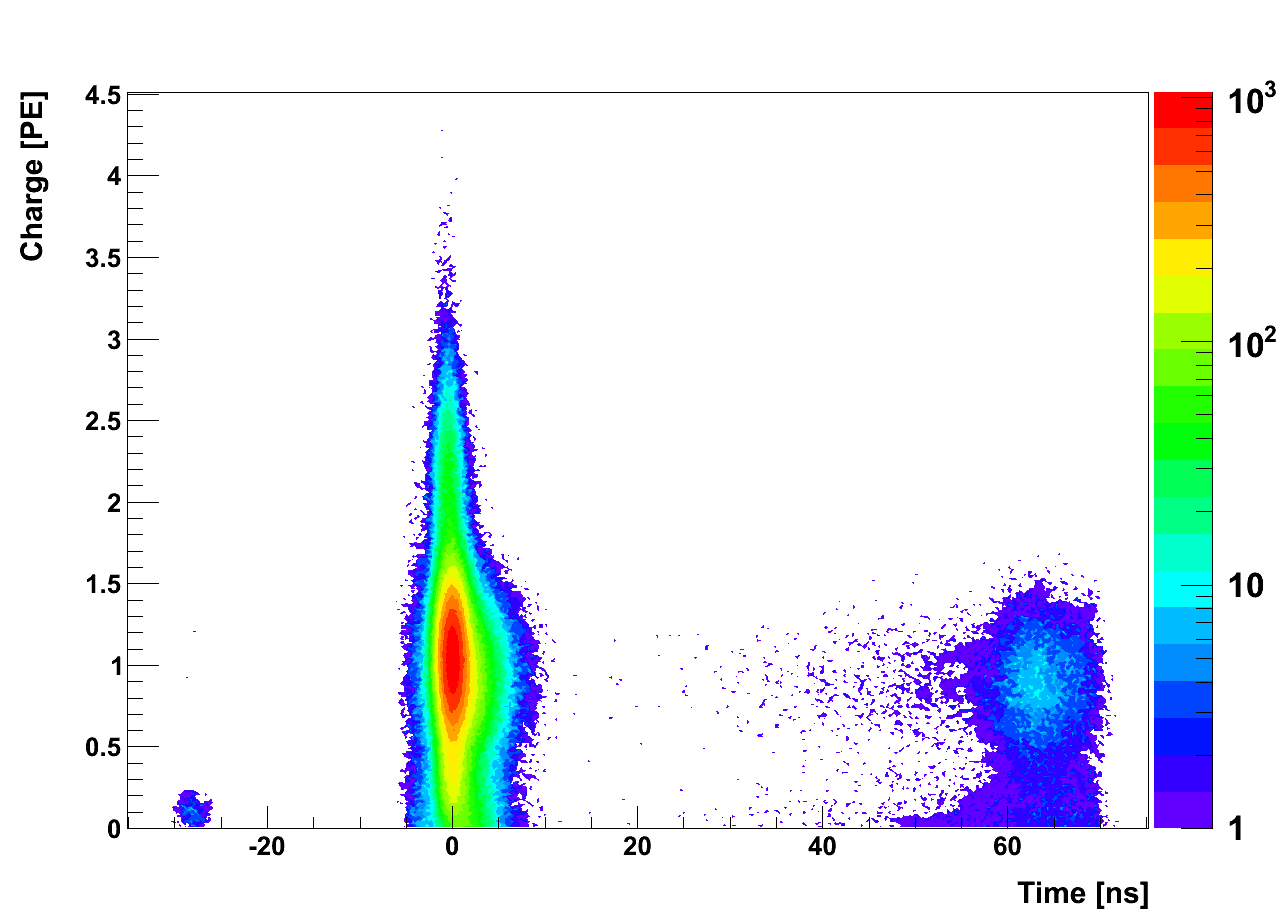}
\caption{Simulated charge versus transit time distribution.}
 \label{fig:Simulated_charge_over_time_spectrum}
\end{figure}

The probabilities of different pulses (presented in the last column of Table \ref{tab:PDFs}) were chosen to be in best agreement with the measured time and charge spectra. The results can be seen in Figure \ref{fig:Simulated_real_time+charge_spectrum}. The charge versus transit time for all simulated events is shown in Figure \ref{fig:Simulated_charge_over_time_spectrum} and can be compared with the measured distribution in Figure \ref{fig:QDC_TDC}.



Also for higher light intensities the simulation yields good results compared to the measured distributions which is illustrated in Figure \ref{fig:Simulated_real_time+charge_spectrum_high_int} for an incident mean number of photoelectrons at the photocathode of $\mu=7.7$. In the real transit time spectrum of figure \ref{fig:Simulated_real_time+charge_spectrum_high_int} the pre-pulse and main peak distribution is wider than in the simulated spectrum, as for this measurement not a laser source but a LED with a larger timing jitter was used. Furthermore, a forward shifting of the main peak of 2$\,$ns can be explained by the fact that larger pulses arrive earlier at TDC threshold. This effect is not caused by the PMT but by applied electronics (the use of a leading edge discriminator) and is therefore not included in the simulation.



\section{Summary}

The single photoelectron response of PMT R7081 was analysed regarding charge and transit time correlations and explained by the underlying physical effects. The results of the measurements were used to build probability density functions for a usage in Monte Carlo simulations which were finally implemented into the Double Chooz detector simulation software.

\section*{Acknowledgements}

We thank the Double Chooz PMT group for the excellent cooperation: Germany (MPIK Heidelberg, RWTH Aachen), Japan (Hiroshima Institute of Technology, Kobe University, Niigata University, Tokyo Institute of Technology, Tokyo Metropolitan University, Tohoku Gakuin University, Tohoku University), Spain (CIEMAT), USA (University of Tennessee).

\end{document}